# HUMIDITY-INDUCED GLASS TRANSITION OF A POLYELECTROLYTE BRUSH CREATES SWITCHABLE FRICTION IN AIR[†]

Stephen Merriman,[1] Saranshu Singla,[1] Ali Dhinojwala[1]*

[1]School of Polymer Science and Polymer Engineering, University of Akron, Akron, OH 44325, USA.
*Corresponding author. Email: ali4@uakron.edu

Electronic supplementary information available upon request.

## ABSTRACT

Polymer brushes have found extensive applications as responsive surfaces, particularly in achieving tunable friction in solvent environments. Despite recent interest in extending this technology to air environments, little is known about the impact of vapor absorption on friction. Considering polyelectrolyte brushes, we report findings that reveal, with increasing relative humidity, a trend of frictional shear forces decaying over two orders of magnitude only after achieving a critical humidity. However, water absorption, structure, and swelling of the brushes followed continuous trends with increasing humidity. In contrast, a humidity-induced glass transition occurred which caused a shift from dry to fluid-like sliding with the water-swollen brush acting as the lubricant. Below the glass transition, friction is large without regard to water concentration; thus, friction transitions sharply. This switching mechanism, shown to be a general property of the hygroscopic polymer, provides new opportunities for switchable friction or other surface actuation from vapor stimuli.

# I. INTRODUCTION

Surfaces consisting of densely end-grafted polymer chains—known as *polymer brushes*—have found a rapidly growing number of applications spanning fields including anti-fouling [1–3], anti-icing [4–6], nanochannel flow regulation [7–9], and ultralow or switchable friction [10–12]. For many such applications, a polymer brush in the presence of a solvent can be made to swell or collapse in response to environmental parameters such as salt concentration, pH, or solvent quality to produce tunable surface structures and properties. While these applications fall within the context of a brush immersed in a liquid solvent, interest has also more recently emerged in considering brushes in air or other vapor environments with applications such as gas sensing [13–15] and vapor-phase separations [16,17], where the applicability of these surfaces is greatly expanded. Evocative of how brush swelling and collapse in liquid solvent can create switchable friction, some studies have also shown that polyelectrolyte (charge-bearing) brushes in air with high humidity can display low friction relative to the dry state [18–20]. While this effect was never explored in detail nor its mechanism, it seems that polymer brushes could then have the potential for use in designing new ultralow or tunable friction surfaces in vapor environments for such applications as catheters [21] or robotics [22], especially if this property could be well understood and tuned for a specific application.

Polyelectrolyte brushes are of particular interest to the field of friction and lubrication since they can absorb and retain water as a lubricant and can exhibit an especially low friction coefficient [23–25]. Previous measurements of water uptake as a function of relative humidity for such brushes typically show absorption curves with a convex shape displaying accelerating water uptake at higher levels of relative humidity [26–29]. Additionally, others have noted strong connections between the swelling state of a brush (swollen/collapsed) [11,12,30] or the amount of unbound water available [31] with a switching friction coefficient for a brush in the presence of liquid water. It may therefore be expected that friction forces should decrease in a similar manner as the water uptake or the swelling of the brush. In this case, friction could be tuned for a specific application by obtaining an appropriate water absorption curve with relative humidity. On the other hand, it is generally viewed that water molecules strongly bound to charged moieties play a critical role in providing ultralow friction for polyelectrolyte brushes [23,24,32–35]. From this view, the population of these strongly bound water molecules within the brush may be more closely linked with changes in friction.

Moreover, some previous work has also pointed out that a glassy polymer brush absorbing solvent vapors will exhibit a plasticization effect whereby the absorbed solvent lowers the glass transition temperature ($T_g$). Past the solvent partial pressure where the $T_g$ has been lowered below room temperature, accelerated absorption of solvent vapor occurs [36,37]. Although demonstrated for organic vapor absorption in nonpolar brushes, polyelectrolyte brushes absorbing water would be expected to show similar effects. Given that many polyelectrolytes or highly hydrophilic polymers exhibit $T_g$ values above room temperature as well as water absorption with increasing humidity, it is reasonable to think that a humidity-induced glass transition could occur within a range of experimentally attainable humidity if the lowered $T_g$ crosses room temperature. Although, prior studies considering polyelectrolyte brush water uptake in humid vapor generally have not considered the possibility of such a transition, possibly due to the continuous nature of the observed water absorption [28,29,38–40]. Yet, a glass transition could itself have implications on the properties of a brush beyond just the water uptake due to the mechanical properties and chain relaxation dynamics that can change rapidly [41,42]. Despite this outstanding knowledge gap and potential applications, no research work to date has examined in detail how the water absorption behavior of a polyelectrolyte brush with humidity affects properties such as friction.

In an effort to explore this topic of considerable interest, the cationic brush poly[(2-methacryloxy)ethyl trimethylammonium] (PMETA) and anionic brush poly(3-sulfopropyl methacrylate) (PSPMA) were chosen as commonly studied polyelectrolyte brush systems that are known to provide ultralow friction in the fully solvated state underwater [11,20,43] but which exist as glassy polymers when dry at room temperature [44–46]. Additionally, monovalent ion exchange was performed with each brush to alter the water absorption and water structure curves within a given brush system. Shear force values during sliding were recorded with incrementally increasing relative humidity and revealed a remarkable decrease in friction that was around two orders of magnitude. This drop occurred within a narrow range of humidity at a critical relative humidity value, before which the shear force was large and unchanging. To elucidate the origin of this transition, water absorption, and swelling profiles with humidity were recorded using attenuated total reflection infrared (ATR-IR) spectroscopy and spectroscopic ellipsometry. In addition, interfacial brush structure and hydration before and during physical contact were probed using sum frequency generation (SFG) spectroscopy. The results revealed that water absorption and structure within the brush bulk and the changing chain structure at the brush surface alone cannot fully explain the trends in friction. However, the transition point coincided with a solvent-induced glass transition as well as a sudden shift in the mechanism of sliding across the brush surface from dry toward a fluid-like state to dramatically lower the shear forces. Based on comparison to spun-cast films, this transition mechanism is shown to be more closely related to the glass transition of these highly hydrophilic polymer films instead of a property of the grafted brush structure. These findings have important implications for the design of surfaces with ultralow or switchable friction in air environments and may also inspire new studies toward how such a transition might be tuned to a desired humidity or extended to other applications of polymer brushes in vapor environments beyond friction.

## II. RESULTS

### A. Dependence of humidity on brush friction

The chemical structures of the cationic brush PMETA and the anionic brush PSPMA chosen for study are shown in Fig. 1A. It is well established that polyelectrolyte brushes can exhibit tunable hydration and structure in response to counterion exchange [47,48], which can lead to changes in observed properties such as friction [11,30]. While these observations are made for brushes in an aqueous solution, it is reasonable to expect that similar tunability could exist for brushes in humid vapors. As a result, counterion exchange was performed for both brush chemistries to include two counterions each – namely $Cl^-$ and $I^-$ for PMETA and $Li^+$ and $K^+$ for PSPMA – to better explore how shifts in hydration may influence friction in a vapor environment. In the sliding geometry used to characterize friction (depicted in Fig. 1A), a polydimethylsiloxane (PDMS) hemispherical lens is brought into contact with the brush-grafted substrate under a chosen humidity condition and made to slide at a fixed velocity of 6 μm/s and a normal load ($F_N$) of 30 mN. Fig. 1B provides an example of the shear force ($F_S$) traces recorded under two extreme relative humidity (%RH) conditions as a function of time for the PMETA-Cl system. When the lens begins sliding at 0% RH, a relatively large shear force reaches equilibrium at around 78 mN. When high humidity vapors (~83% RH) were suddenly introduced, after some delay in time corresponding to the time required to displace the dry vapor in the chamber, the shear force reduced dramatically to around 2–5 mN. This change was reversible and could be brought back to 78 mN by reversing the humidity to 0%. This drop in friction between low and high humidity is consistent with that observed from previous reports [18–20]. However, this observation alone does not allow us to readily comment on the nature of the drop nor its connection to other properties of the brush.

To obtain greater detail, we show the complete trends in the shear forces as a function of %RH in Fig. 1C for all four brush systems. The data are provided in terms of the equilibrium shear force achieved during sliding at each condition with $F_N = 30$ mN, shown on a log scale (linear scale provided in Fig. S1 in supplementary material). If instead considered as a friction coefficient, the slope of $F_S$ versus $F_N$ at the extreme conditions corresponds to a $\mu_k$ value of 1.52 at 0% RH and 0.0084 at 95% RH (Fig. S2 of supplementary material). In all cases, no measurable change in shear force was observed for the entire range of humidity below around 50% RH. However, in the higher humidity regime, each brush system showed a sharp transition point after which the shear forces decayed rapidly. In this transition region, slight changes in %RH generated large changes in shear forces which led to relatively large error bars. As a control to ensure that the sharp friction change was not due to humidity interfering with the instrument or a result of water condensation, the shear forces for hydrophobic and hydrophilic self-assembled monolayers was also recorded (Fig. S3 of supplementary material); as no large measurable change with humidity was noted, the measured trends for these brush systems were thus confirmed. The inset of Fig. 1C shows the zoomed-in region between 60% and 100% RH to better visualize the differences between systems, separated by polymer type (PMETA or PSPMA). PMETA-I shows a transition point at a higher %RH compared to PMETA-Cl, and the shear forces did not decay to values as low as those of PMETA-Cl. The trends for PSPMA-Li and PSPMA-K were comparable, but PSPMA-K displayed a shear force drop slightly shifted to a higher %RH. The shear force trends for polymers PMETA-Cl and PSPMA-Li were indistinguishable. Overall, the critical %RH where we observed a shear force drop varied between around 55%–75% RH depending on the system. While the magnitudes of the shear forces at the upper humidity limit are very low for most systems (~0.9 mN), it is worth noting that even lower values below the detection limit of our tribometer were recorded for sliding underwater (Fig. S4 of supplementary material) and therefore could be some orders of magnitude smaller than in humid vapor. The nature of this large drop in friction, which is highly discontinuous, has not been previously reported and raises important questions about the mechanism behind such a transition.

To explore further, we measured the variation in the shear force with sliding velocity to investigate what dissipation processes account for the observed forces (Fig. 1D). Different dissipation processes provide shear forces which vary differently as a function of shear velocity and, thus, can be a useful tool for elucidating such processes [49]. We recorded the PMETA-Cl brush shear forces as a representative system at both the high (95%) and low (0%) humidity regimes as a function of sliding velocity. At 0% RH, no significant dependence of sliding velocity on shear force was found for the velocity range studied. However, at 95% RH, a significant positive trend scaling shear force with sliding velocity ($v_s$) was noted as $F_S \sim v_s^{0.54}$. This shift in velocity dependence indicates a fundamental shift in the dissipation pathways responsible for the observed shear forces as described by the classical Stribeck curve from dry-type to hydrodynamic-type lubrication [49] since each dissipation mode scales differently with velocity, hinting towards an interfacial change which we consider in greater detail in the following sections.

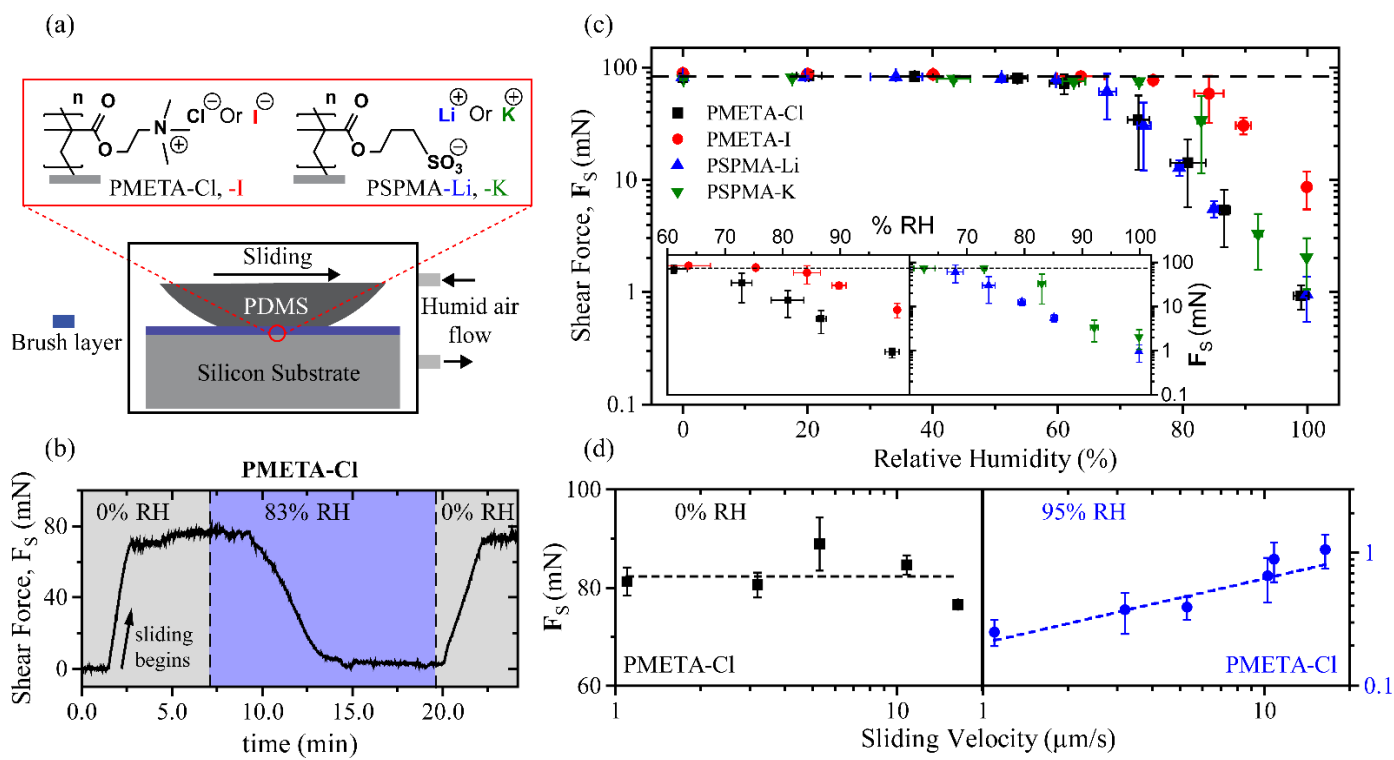

Fig. 1. Dependence of brush friction with increasing relative humidity. (a) Chemical structures for the brushes and counterions studied and illustration of the PDMS–brush contact within a humidity chamber used to measure friction. (b) Response of measured shear force upon sliding for PMETA-Cl surface and suddenly changing humidity between 0% (*grey regions*) and ~83% (*blue region*). Initial sliding begins at $t = $ ~2 min. (c) Trends in equilibrium shear force values (log scale) with incrementally increasing humidity for each brush and ion system. The inset shows the zoomed-in region between 60 and 100% RH for the two separate polymers to highlight differences. (d) Shear force as a function of sliding velocity for the PMETA-Cl brush at 0% RH (*black squares*, linear-log plot) and ~95% RH (*blue circles*, log-log plot). Dotted lines represent a fit to the mean for 0% RH to indicate the lack of a significant dependence of shear force with sliding velocity, and a power law fit for 99% RH to indicate a significant positive dependence of shear force with sliding velocity. Statistical significance was determined using an ANOVA test for both fitted power law and linear functions with a significance cutoff of $p = 0.05$. The shift in velocity-dependent behavior with humidity indicates a shift in the lubrication mechanism.

## B. Water absorption, structure, and swelling of the polyelectrolyte brush

The ultralow friction provided by polyelectrolyte brushes and similar highly hydrophilic materials in aqueous environments has usually been described as being controlled by water molecules tightly bound to charged groups of the polymer [23,24,32–35], the presence of large quantities of absorbed water that are not easily removed under load and that act as a lubricant [31], or a consequence of the highly extended brush structure in the swollen state [11,30]. It may therefore be anticipated that the shear forces should closely follow a quantity that changes with humidity, such as water content, water structure, or swelling ratio. To directly address this, ATR-IR spectra of the brush layers were recorded as a function of humidity using the geometry shown in Fig. 2A. A calculation of the penetration depth of the IR beam into the brush sample (in supplementary Fig. S5 and supplementary text) suggests that the data obtained from such spectra should be representative of the entire length of the brush film at every humidity condition. The collected spectra under 0%, 45%, and 90% RH for each brush system are presented in Fig. 2B. Each set of spectra show similar absorbance signatures including a broad OH stretching mode ($\nu$OH) around 3100–3700 cm$^{-1}$ and a weaker HOH bending mode ($\delta$OH) at around 1640 cm$^{-1}$, which are assigned to absorbed water molecules. Additional peaks assigned to the brushes themselves were observed including aliphatic CH stretching modes present around the 2750–3000 cm$^{-1}$ region and a sharp carbonyl stretching mode at around 1720 cm$^{-1}$ ($\nu$CO). The spectra for PMETA-Cl and PSPMA-K are in good agreement with those previously reported by Lin *et al.* [29]. With increasing humidity, the peaks assigned to water molecules increase whereas those assigned to the brush remain of similar intensity. To quantify these changes, a ratio between the height of the water bending ($\delta$OH) and carbonyl stretching ($\nu$CO) peaks was calculated from each spectrum (Fig. 2C) and taken as a quantity intrinsically related to the volume fraction of water within the brush based on a Beer–Lambert law approximation. The results indicate a gradually increasing trend in water uptake with an increasing slope from low to high humidity comparable to that observed by others for similar systems [26–29,36]. Of note, each system also shows a steep linear rise in water content at very low humidity (0–10% RH) just like that observed by Laschitsch *et al.,* who described it as a Henry sorption regime or as being related to a void or "hole filling" mechanism [36]. Between brush systems, we observed similar trends, but with vertical shifts with PSPMA-Li showing the greatest water content and PMETA-I showing the smallest water content. For a given polymer, the smaller counterions provided greater water absorption; this result is in agreement with a comparison to the hydration state and brush structure underwater for these ions as well as with trends described by the Hofmeister ion series [47,48]. While each brush certainly absorbs water with increasing humidity, this finding does not support a direct correlation between water absorption and a reduction in friction. Significant increases in water content were observed for all systems up to 50% RH despite no change in friction. Furthermore, trends in water absorption between systems suggest that PSPMA-Li should have the lowest friction, or the earliest friction drop with increasing humidity; however, we did not find this to be the case.

To investigate further, we quantified the shifting water structure within the brush from the ATR-IR spectra. In Fig. 2B, dashed vertical lines are positioned as a visual guide to demonstrate how the center of the OH stretching peak shifts to lower wavenumbers with increasing humidity to different extents depending on the system. Others have used the shift in the position of this peak to describe the overall strength of the hydrogen bonding network of the probed water molecules, where an absorbance near 3200 cm$^{-1}$ corresponds to a strongly-bound "ice-like" water structure and an absorbance near 3600 cm$^{-1}$ corresponds to a more weakly bound or "free" water structure [4,50–52]. While Bonn *et al.* have raised concerns about this specific interpretation of water signatures [52–54], we believe that such concerns do not prohibit the analysis of shifts for a given system between spectra. As a simple way to quantify this shift, a ratio of the area between the regions 3265–3060 cm$^{-1}$ and 3745–3060 cm$^{-1}$ was calculated as a percentage of the total water within the brush that adopts a strongly bound ice-like structure. The calculations for each system as a function of humidity are presented in Fig. 2D. Each system shows a water structure trending toward the ice-like state with some vertical shifts from one another. Excluding the region below 10% RH, the PMETA brushes with both counterions show water structures that are changing linearly across all %RH values. For PSPMA, slope changes were observed at 30% RH and 50% RH for Li$^+$ and K$^+$, respectively. These observations, although they are interesting and point toward key differences in how water can arrange and interact within these partially hydrated brushes, seem to bear no direct connection to the measured trends in friction, as no clues are given as to why friction should drop at the observed critical humidities or why relative shifts in critical humidity are observed for the two brush systems.

As a final test of parameters that have been previously suggested to play a key role in driving brush friction, we measured brush thickness and refractive index using spectroscopic ellipsometry by fitting raw Ψ and Δ values to a model with negligible dependence of wavelength on refractive index (Fig. S6 of supplementary material). The resulting values are shown in Fig. 3A for PMETA-Cl with thickness reported as a swelling ratio $d/d_0$ with a dry brush thickness measured for each sample of ~110 nm. Further, refractive indices and swelling ratio values were converted into water volume fraction within the brush using an effective medium approximation and volume balance, respectively, as outlined in the supplementary text. The results of these calculations are shown in Fig. 3B. Since ATR-IR measurements confirm that water exists within the brush even at 0% RH while the calculations assume a water fraction of 0 at a humidity of 0%, the values should only be taken as values relative to the water fraction at 0% RH. In agreement with the water uptake measured using ATR-IR, these results point toward a scenario where the brush absorbs water and swells across the full range of humidity, including the low humidity regime where no changes in friction were observed. Thus, thickness change and water absorption alone cannot allow us to understand the mechanism behind the observed transition in friction. However, we did observe a change in slope (as noted in Fig. 3B), and this will be discussed in the following section.

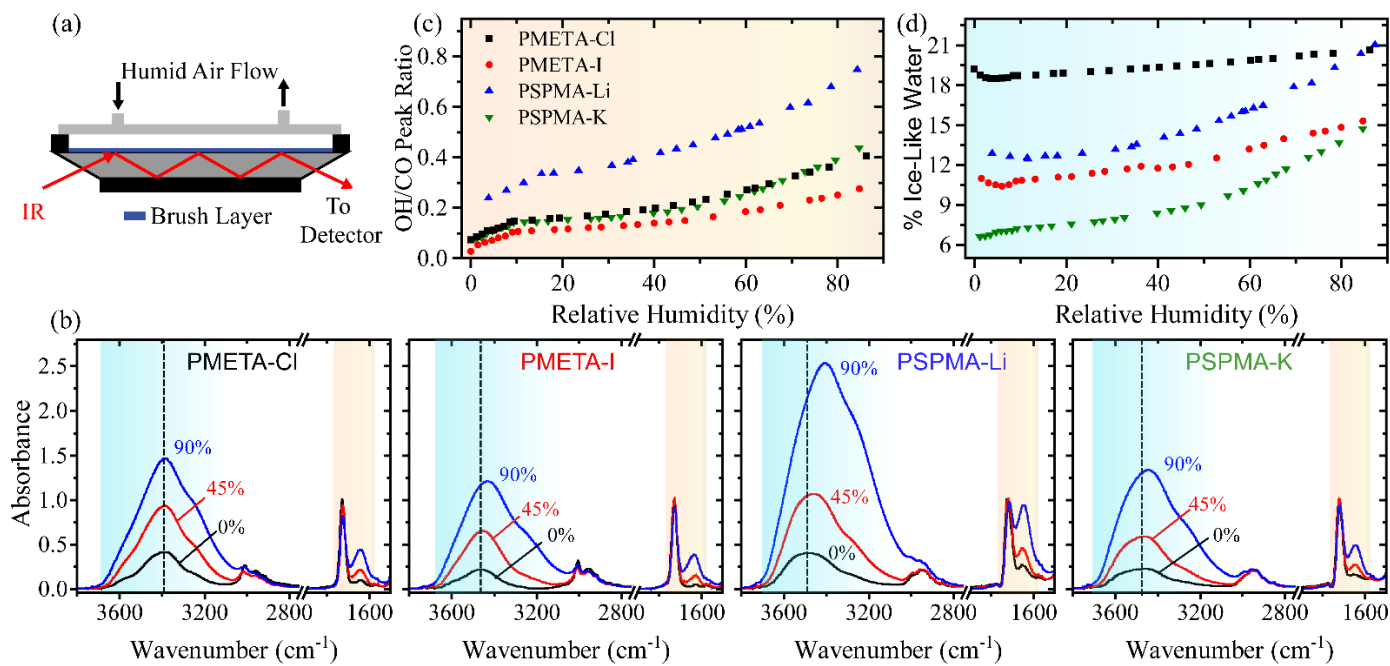

Fig. 2. Brush bulk probed using ATR-IR spectroscopy. (a) Schematic of ATR-IR spectra collection geometry. The number of bounces is not shown to scale. (b) The collected ATR-IR spectra for each brush and ion system at 0, 45, and 90% RH. Dashed vertical lines are presented as a visual guide for observing peak shifts. Absorbance for each set of spectra is normalized by the carbonyl peak ($\nu$CO) intensity at 0% RH for a more direct comparison of absorbance between samples. (c) The trends in water HOH bending ($\delta$OH)/brush carbonyl ($\nu$CO) peak height ratios calculated from ATR-IR spectra, which is a quantity related to the fraction of water in the brush bulk, as a function of humidity for each brush/ion system. (D) The trends in relative percentage of ice-like water vs. %RH. We define the percentage of ice-like water as the area ratio between the region 3265–3060 cm$^{-1}$ (ice-like shoulder) and the entire water stretching region ($\nu$OH) 3745–3060 cm$^{-1}$.

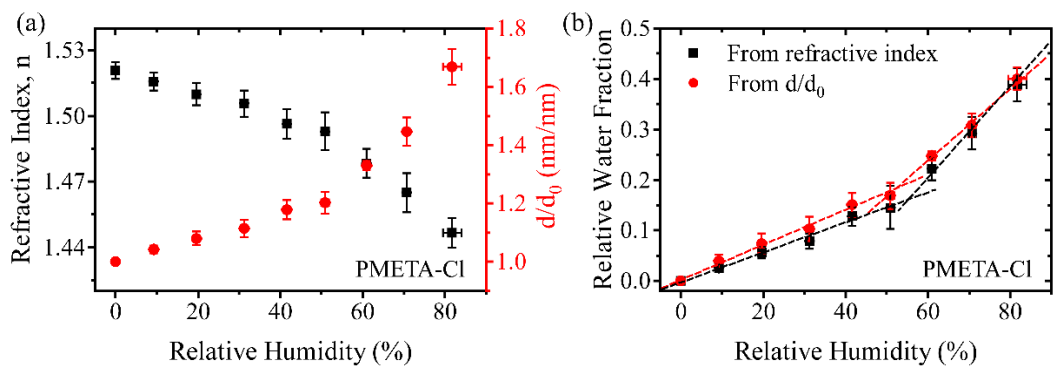

Fig. 3. Brush thickness, refractive index, and water content from ellipsometry. (a) Refractive index (n) and swelling ratio (d/d$_0$) with d$_0$ ~ 110 nm for PMETA-Cl brush system as a function of humidity. (b) Relative volume fraction of water within the brush layer calculated using both refractive index values (*black*) and swelling ratio values (*red*). A change in the rate of water uptake is shown for relative humidity beyond ~55%, as illustrated by the dotted linear fits.

### C. Humidity-induced glass transition

While brush swelling does not directly compare to the friction trends, it is clear from the dataset in Fig. 3B that a two-regime swelling behavior exists with a different slope above a critical humidity of around 55% RH with both swelling ratio and refractive index. We observed similar slope changes in the water content measured with ATR-IR, as shown in Fig. 2C. Given the reported values of around 180 ºC [44,45] and 187 ºC [46] for the respective $T_g$ values of bulk PMETA and PSPMA in dry conditions and the expected plasticization effect of absorbed water to lower $T_g$ [36,42], it is possible that these slope changes result from a humidity-induced glass transition. Engkagul *et al.* noted the apparent presence of a glass transition for bulk-prepared PMETA at some point below 90% RH [44]. Since PMETA and PSPMA share similar $T_g$ values in the dry state and exhibit similar water uptake, it is reasonable to expect a humidity-induced glass transition for both systems within this range of humidity. In addition, surface grafted brushes, as compared to the bulk polymer, are expected to show similar $T_g$ values for brush thicknesses greater than 40 nm [55]. Galvin and Genzer have also noted the possible existence of such a transition from brush water absorption trends [27].

To investigate potential changes in modulus due to a glass transition, we performed nanoindentation measurements on thicker 350-nm PMETA-Cl brushes as a function of humidity. Fig. 4A shows the raw loading and unloading curves at different humidity conditions with a 50 nm fixed displacement depth into the brush. The peak indentation force achieved reduces with increasing humidity and

Page **8** of 24

accelerates as humidity is further increased. Beyond around 63% RH, we were no longer able to obtain a proper indentation curve as the forces were below the noise and detection threshold of around 1 μN. From the unloading curves, values of reduced modulus were obtained and are presented in Fig. 4B. At low humidity, we note that the measured modulus values of around 20 GPa are very large compared to typical values for glassy polymers in the 1-4 GPa range. Although we purposely chose thicker brushes for these measurements, this is still likely due to the influence of the silicon substrate which has a modulus of around 170 GPa [56]. For thin film nanoindentation experiments, it has been reported that the substrate modulus has a significant influence on measured results even when the indentation depth is a small fraction of the film thickness [57]. At the highest humidity measurable of 63%, the modulus value dropped to 350 MPa with a range as low as 150 MPa, representing a two-order of magnitude decrease. While the brush is expected to increase in thickness around 20% (to 420 nm) by this humidity (Fig. 3a) which could itself reduce the measured modulus, this effect alone cannot explain this magnitude of decrease. In general, micron-scale film thicknesses are required to neglect substrate effects; swelling by an additional 70 nm should not greatly impact the measured modulus. Therefore, this trend with such a large decrease in modulus is indicative of a humidity-induced glass transition, with further decreases to near-zero expected at larger humidity if such moduli could be measured through nanoindentation [58]. For comparison, Fig. 4B also highlights a shaded region where the friction transition was measured for the same system in Fig. 1C. While we measure changes in modulus across the entire range of humidity, the concurrence of the friction transition region where the modulus is most rapidly decreasing to unmeasurable values indeed suggests a plausible connection between these two phenomena.

An additional example demonstrating the presence of the humidity-induced glass transition is presented in Fig. S7 in the supplementary material. Here, a simple mechanical creep test was performed hanging a sheet of PMETA-Cl bulk-prepared polymer while supporting a 20-g weight in a chamber with relative humidity slowly increased at a rate of 0.9 %RH/min and monitoring the changes in dimensions. Up to 50% RH, we observed no changes to the polymer sheet. By 55% RH, we observed a small but detectable extension of the sheet, and only a small further increase in humidity to 57.4% RH resulted in large extension and necking of the sheet. The extension ratio $L/L_0$ for the PMETA-Cl sheet in each photo, determined using the edges of the sheet and anchoring clip, is provided. The onset of extension is first noted at around 48% RH, after which extension accelerates rapidly until the polymer sheet breaks at around 58% RH. This onset of extension is indicative of a glass transition for the polymer occurs within this measured humidity range, which very nearly coincides with both the friction transition in Fig. 1C and the swelling slope change in Fig. 3B.

Given that a glass transition was shown to occur, an analysis of the glass transition relative humidity between all four brush/counterion systems is justified; the analysis was conducted from water absorption data measured by ATR-IR using the previously proposed methodology [36,59]. Examples of the process are provided in Fig. 4C for PMETA-Cl and PMETA-I, where linear fits were calculated within the linear regions at high and low humidity and where the crossover point was taken as the glass transition relative humidity, $RH_g$. Examples of PSPMA-Li/K are also provided in Fig. S8 in the supplementary material. We note that the glass transition occurs over a range with width spanning at least 30% RH as indicated by the slope transition region. The point taken as $RH_g$ is somewhat arbitrary in nature but comes from an analysis that minimizes ambiguity and subjectivity. The results from this procedure are presented in Fig. 4D. In this figure, solid bars represent the $RH_g$ as determined from the ATR-IR analysis, while dashed bars represent the friction transition points for comparison. Letters A and B indicate groups having significantly different $RH_g$ values based on statistical analysis.

PMETA-I shows a significantly higher $RH_g$ value of 60.4%, while the remaining three systems show similar $RH_g$ values ranging between 42.8% and 48.8%. By comparing these $RH_g$ values to the %RH values where the drop in friction occurred (dashed bars), the higher %RH friction transition of PMETA-I and relatively similar %RH transitions of PMETA-Cl and PSPMA-Li are explained well by a shift in $RH_g$. Following this argument, the friction data suggests that PSPMA-K should have an $RH_g$ value somewhere between those of PMETA-Cl and PMETAC-I; however, this could not be proven within the limit of statistical precision of these experiments. However, in sharp contrast to each of the previously observed trends in water content, structure, and swelling, these shifts in $RH_g$ trend well with the friction transition shifts. It follows that as humidity increases, $T_g$ is lowered; at the point where $T_g$ crosses room temperature ($RH_g$), we observe the sharp transition in friction. By shifting $RH_g$, the friction transition point is shifted.

We point out that these $RH_g$ values, though they trend well with the friction transition points, have distinctly lower values. However, as noted, a humidity-induced glass transition occurs across a broad range of %RH values with the mechanical properties of storage modulus, loss modulus, and loss tangent following different trends at different stages of the glass transition [58]. Although we do not know at what point we should expect a transition in friction, we find that a more quantitative connection with the friction drop occurs with the endpoint of the glass transition region as presented in Fig. 4C. One possible reason for this comes from the expectation that past this point, the brush enters the Flory-Huggins regime with the polymer chains behaving more like a polymer solution [36]. This also agrees well with the observations of the friction transition coinciding closely to that observed from brush swelling acceleration (Fig. 3) and the most rapid drop in modulus (Fig. 4B). Since each of these observations fundamentally relates to the glass transition process of the polymer and the shifts in $RH_g$ agree well with shifts in the friction transition point, we conclude that the humidity-induced glass transition is driving the measured friction transition.

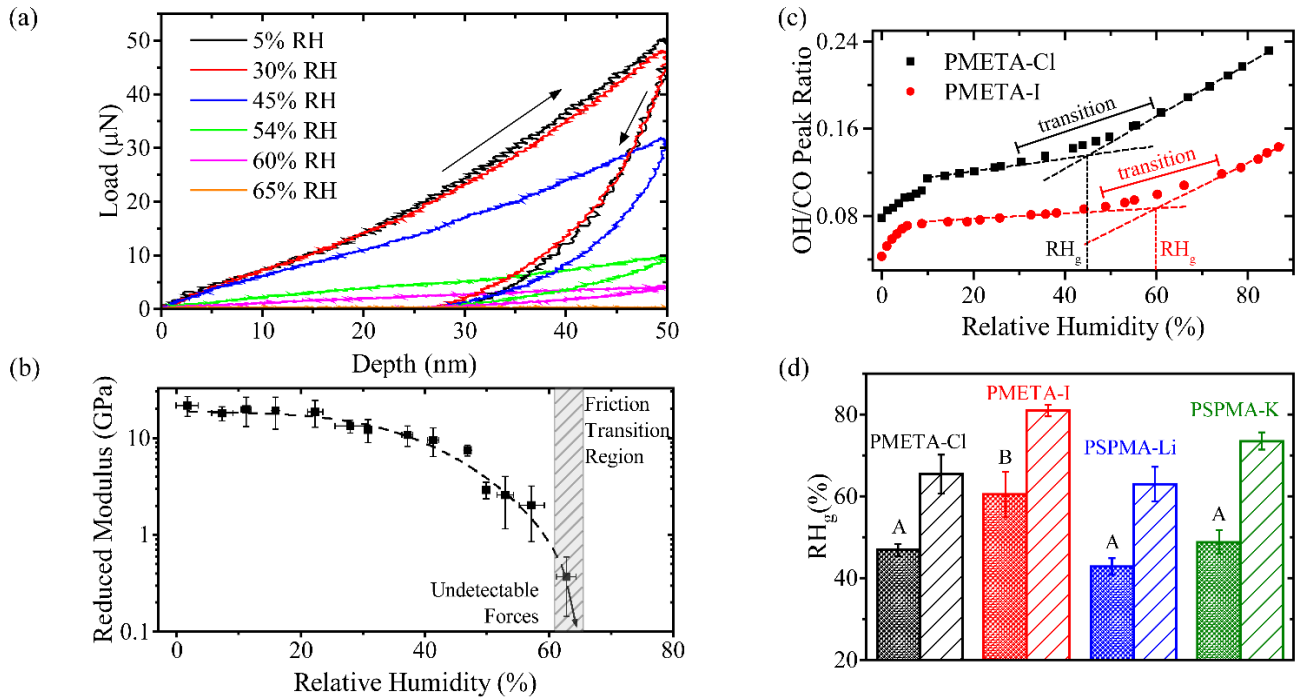

Fig. 4. Probing the presence of a humidity-induced glass transition humidity, $RH_g$. (a) Raw loading and unloading nanoindentation curves for the PMETA-Cl at different humidity levels. Arrows indicate the direction of loading/unloading. (b) Reduced modulus extracted from the nanoindentation unloading curves plotted as a function of relative humidity on a linear-log scale. As noted, past 63% RH the indentation forces become undetectably low. The shaded region is provided as a reference for where the friction transition occurred from Fig. 1. (c) Determination of glass transition humidity, $RH_g$, from ATR-IR OH/CO peak ratios like those shown in Fig. 2B for the example cases of PMETA-Cl and PMETA-I (the plots for the other polymers are provided in supplementary Fig. S8). (d) Glass transition humidity values $RH_g$ for each brush/ion system, where the letters A and B indicate groups with a statistically significant difference in $RH_g$ values. Statistical significance was determined using a Tukey–Kramer honest significant difference test on mean pairs using a significance cutoff of $p = 0.05$. Solid bars were determined from the midpoint of the region shown in Fig. 4C from ATR-IR data, while dashed bars are the transitions observed from the friction data in Fig. 1C.

### D. Probing the brush contact interface

With the presence of a humidity-induced glass transition demonstrated to coincide with a dramatic drop in friction for these brush systems, our attention turns towards understanding the mechanism for the sudden change in friction. For example, Figs. 2 and 3 demonstrate that the brushes absorb water and swell in the bulk (along with modulus changes in Fig. 4B) even below the glass transition when the shear forces do not change. It is not clear from this data alone why the glassy brush should have shear forces that are entirely independent of bulk water absorption. However, it is reasonable to suspect that the glass transition is driving an interfacial change that occurs abruptly at the glass transition and that does not depend on water absorption within the brush bulk. To address this, we collected surface-sensitive SFG vibrational spectra of the brush surface. With SFG spectroscopy, no signals are generated from centrosymmetric bulk media; thus, the resulting spectra pertain only to the interface [60,61]. SFG therefore allows us to comment on interfacial brush structure and hydration state by analyzing the observed hydrocarbon signatures (from the brush) and hydroxyl signatures (from surface-bound water) along with their intensities and wavenumber positions.

In these experiments, we grafted a PMETA-Cl brush to a sapphire prism and collected spectra under different levels of relative humidity before and after bringing a PDMS lens into contact with the brush (Fig. 5A). Figs. 5B and 5C show spectra in PPP polarization (P-polarized SFG, visible, and IR beams) at 0% and 85% RH before contact with the PDMS lens in the hydrocarbon and hydroxyl regions, respectively. Hydrocarbon peaks observed include 2820, 2875, and 2920 cm$^{-1}$, which were assigned to the symmetric $CH_2$, symmetric $CH_3$, and $CH_3$ Fermi resonance stretching vibrations from the brush [29,62,63]. These peaks appear to proportionally decrease in intensity with increasing relative humidity. Peaks in the hydroxyl region are observed primarily around 3600 and 3700 cm$^{-1}$, which have been assigned to weakly interacting and non-interacting OH vibrations, respectively [64,65]. Calculations using Fresnel factors suggest that these spectra should be dominated by the brush–air interface rather than the brush–sapphire interface (Fig. S10 of supplementary material), thus assigning these peaks to very weakly hydrogen-bonded water on the brush surface. However, the contribution from the surface hydroxyl groups on the underlying sapphire substrate that show similar signatures at these wavenumbers cannot be ruled out. Between 3000 and 3400 cm$^{-1}$, a relatively weak broad signature exists that does not change considerably with humidity, indicating the presence of water with a liquid-like structure [66]. While signal intensity appears much larger for the 3500–3700 cm$^{-1}$ than the 3000–3500 cm$^{-1}$ region, this is attributed to the fact that the SFG signals in total internal reflection geometry are higher in this region due to wavenumber-dependent Fresnel factors [67].

To better track intensity changes as a function of relative humidity, we fixed the IR wavenumber at values of 2920, 3600, and 3700 cm$^{-1}$ while incrementally increasing humidity and tracking the resulting SFG intensity as shown in Fig. 5D. Each wavenumber shows similar decreasing trends with humidity that accelerate after the glass transition comparable to the trends in water uptake and the swelling of the brush. Changing Fresnel factors cannot account for these intensity changes, as such calculations instead predict increasing intensity (Fig. S10 of supplementary material). Although, SFG signal intensity is sensitive to the degree of average orientation of vibrations, and thus these decreasing intensities can be attributed to an increase in the time-averaged disorder or mobility of these functional groups as the chains absorb water and rearrange or become more mobile. This effect has been previously observed for the solvation of various monolayers [68,69] and can also be inferred from the fact that the reported PMETA-Cl spectra under 0% RH show relatively intense peaks as compared to those previously reported for PMETA-Cl fully solvated by water [62] and decreases in intensity with a trend matching that of the measured water content in the bulk of the brush. Interestingly, the SFG intensity decreases across the entire range of humidity but accelerates past the glass transition point. Since changes were observed even in the low humidity region, no direct connection can be made between the changes in structure and/or mobility of surface chains and the friction trends. However, the onset of the glass transition can be clearly identified on the brush surface from this dataset, indicating the effect of the glass transition point on the increase in interfacial chain reorganization or mobility.

The interfacial contact region between PDMS and the brush was also directly probed under controlled humidity. Figs. 5E and 5F show the equivalent SFG spectra after bringing the PDMS lens in contact with the PMETA-Cl brush for cases of 0%, 40%, and 95% RH. These spectra show similar signatures and trends to those obtained before contact with the addition of peaks having negative amplitudes (suggesting opposite orientation relative to neighboring brush peaks) at around 2880 and 2950 cm$^{-1}$, which is assigned to the symmetric and asymmetric CH$_3$ vibrations from the PDMS [66]. Additionally, the broad water signature between 3000 and 3400 cm$^{-1}$ is reduced for all cases except 95% RH. The brush in contact with PDMS adopts a similar structure and mobility as before contact across the entire range of humidity but tends to show water of liquid-like structure primarily only at higher humidity. Peaks near 3600 and 3700 cm$^{-1}$ are present in contact for all three values of %RH, but these peaks are more difficult to comment on since their origin is not fully clear. The same qualitative trends were also observed with SSP polarization (S-polarized SFG, visible; P-polarized IR beams) before and after contact (Fig. S9 of supplementary material), suggesting the trend is not specific to an orientation axis probed by these two polarization combinations. We also note that the absence of a water signature from a spectrum does not guarantee the absence of water on the brush surface, since water molecules adopting a random orientation would provide no SFG signal. Still, the presence of a water signature for in-contact spectra only for 95% RH could be significant, as it points toward a shift in the ability of the interfacial region to retain water within the brush.

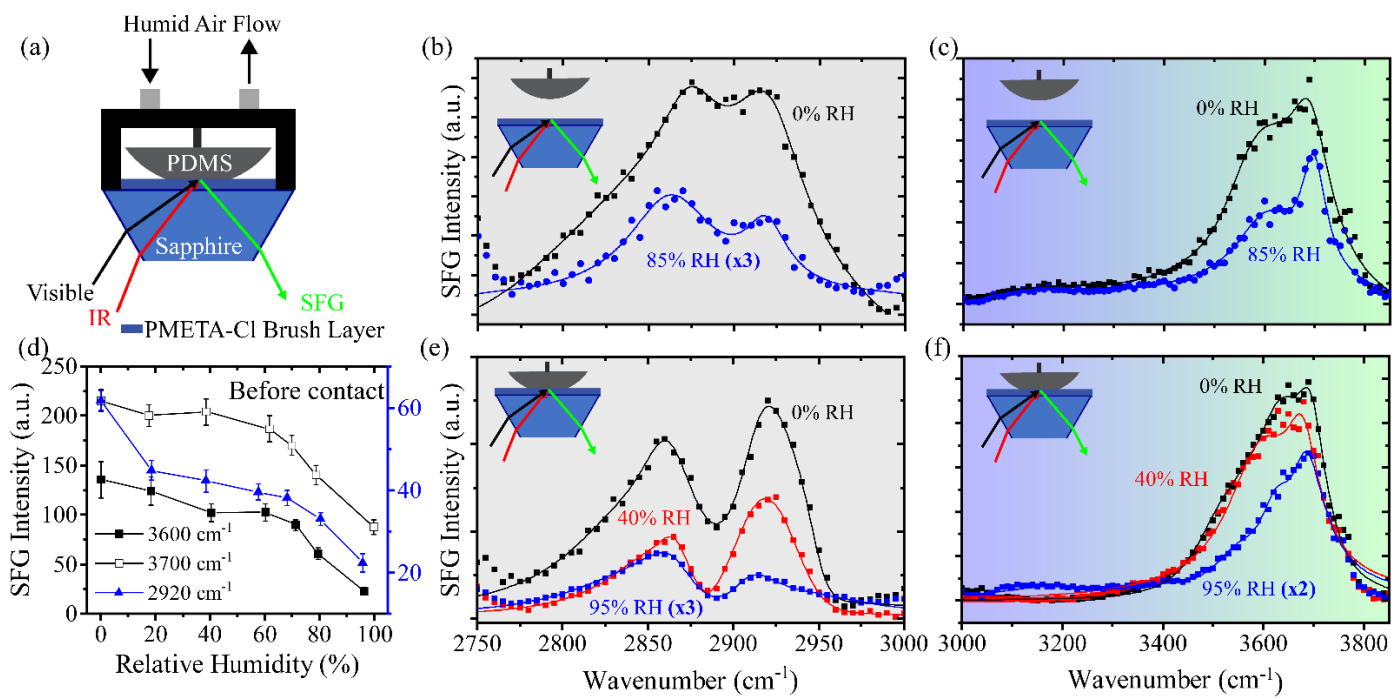

Fig. 5. SFG spectra of brush interface in and out of contact. (a) Diagram showing the SFG spectra collection geometry. (b) Hydrocarbon region (2750–3000 cm$^{-1}$) SFG spectra in PPP polarization at the PMETA–air interface under 0% humidity (*black*) and 85% humidity (*blue*). (c) Hydroxyl region (3000–3850 cm$^{-1}$) SFG spectra in PPP polarization at the PMETA–air interface under 0% humidity (*black*) and 85% humidity (*blue*). (d) Trend in SFG intensity at the PMETA–air interface as a function of humidity for 2920, 3600, and 3700 cm$^{-1}$ IR wavenumbers. (e) Hydrocarbon region (2750–3000 cm$^{-1}$) SFG spectra in PPP polarization at the PMETA–PDMS contact interface under 0% humidity (*black*) and 85% humidity (*blue*). (f) Hydroxyl region (3000–3850 cm$^{-1}$) SFG spectra in PPP polarization at the PMETA-PDMS contact interface under 0% humidity (*black*) and 85% humidity (*blue*).

### E. Extension of friction transition to similar systems

We finally address the question about the broad applicability of these results to other systems. Since the SFG results point toward a drier contact at low humidity, it is possible that the PDMS lens could be a critical component to achieve the abrupt friction transition. We address this question by measuring the shear forces for a brush-grafted PDMS lens sliding on a brush-grafted silicon wafer as a function of relative humidity (Fig. S11 of supplementary material). Compared to the bare PDMS–brush friction shown in Fig. 1C, the region below ~55–60% RH showed slightly lower shear forces and exhibits some dependence on %RH. However, at higher humidity, the shear forces decreased in a sharp similar manner to the case for a bare PDMS–brush. While the modification of the PDMS lens with the PMETA-Cl brush influences the low humidity region, possibly by disrupting the dry contact observed for the bare PDMS, the shear forces for such a partially lubricated solid–solid contact is still clearly much larger than when the brush transitions above $RH_g$. As a result, the same switching friction behavior occurs even for brush–brush sliding.

The second question we address is the importance of a polymer brush in this friction transition or if this is a general property of the polymer–solvent vapor system, as the glass transition phenomenon would suggest. We tested this hypothesis by measuring shear forces for a bare PDMS lens sliding against a spun-cast polymer film of the same chemistry and thickness as the brush layer (shown in Fig. S12 of supplementary material). While spun-cast films are expected to show different water absorption as compared to the grafted brush, the reported differences are small and may not largely affect the glass transition humidity [27,40]. In this case, we observe the same high-to-low shear force transition past $RH_g$ just like that for the grafted brush, which occurs at a similar %RH. This observation confirms that the friction trend is closely related to the humidity-induced glass transition rather than anything specific about the brush architecture, providing support for the argument that this may be a much more general phenomenon. However, the brush-layered films remain intact after sliding at high %RH compared to the spun-cast films, resulting in a much more robust system for industrial applications.

## III. DISCUSSION

The results of this study indicate a sudden drop in the frictional shear forces by over two orders of magnitude for PMETA and PSPMA brush-grafted surfaces at a critical %RH. While the friction transition point coincides with accelerating water absorption and swelling, substantial absorption still occurs below the transition point, despite no changes in friction. Moreover, no correlation exists between how much water the brush absorbed or the brush structure and the value of relative humidity where we observed an onset of friction reduction. Instead, the friction drop coincides well with a humidity-induced glass transition, and the shifts in this transition generated shifts in the shear force curves. This transition also influenced the velocity-dependence of the shear forces and retention of interfacial water at interface of the PDMS and the brush-coated sapphire interface as observed from the SFG spectra. The velocity-dependent friction behavior at low and high humidity (Fig. 1D) can be explained using the Stribeck curve [49]. In the low humidity regime, we measured the shear forces to be independent of sliding velocity. This is indicative of dry-type lubrication, where solid–solid contact occurs and where repeated making and breaking of bonds at the sliding interface give rise to relatively large shear forces. At the high humidity regime, shear force had a positive scaling relationship with sliding velocity. This trend provides evidence of fluid or hydrodynamic-type lubrication, where a fluid film exists between the sliding surfaces and where friction is a purely viscous process that depends on shear rate [35,49]. Collectively, these experiments point toward a sudden shift from a solid to a fluid-like interface as the system crosses the humidity-induced glass transition.

In Fig. 6, we illustrate our proposed model based on the collected data. Although we focus on the case of a brush, the proposed model is not contingent on surface-grafted chains, as the same mechanism holds for spun-cast films. At humidity below $RH_g$, the glassy polymer brush acts as a solid layer that absorbs water and swells with increasing humidity. However, friction is high and humidity-independent since there is solid–solid sliding with a largely dry contact interface. Slip occurs at this dry boundary between the sliding solid surfaces, and we measured large shear forces that were independent of both the sliding velocity and the amount of water present within the brush layer bulk. At humidity above $RH_g$, the humidity-induced glass transition can significantly reduce the relaxation times of the polymer brush chains, which permits the flow of solvent through the film, allowing it to behave more like a fluid than a solid. When sheared, there is a mechanism shift, and the viscous dissipation from this fluid film is responsible for the shear forces, giving rise to the positive velocity dependence and a considerable reduction in magnitude as compared to the dry state. Since the glass transition occurs across a relatively narrow range of relative humidity, friction values transition sharply. We note that the thickness of such a fluid film that experiences viscous shear cannot be determined from these experiments alone.

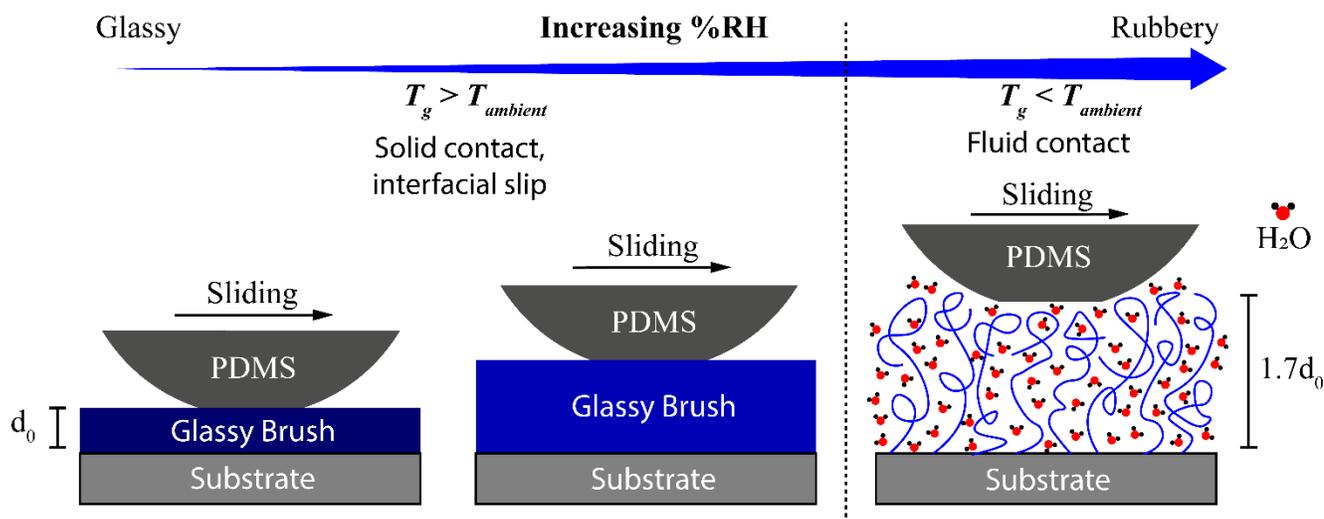

Fig. 6. Molecular illustration of brush humidity response and implications for friction. Before the occurrence of $RH_g$, the polymer is in a glassy state; even though there is water absorption and swelling with increasing humidity, the interface remains dry and the polymer film remains rigid, inhibiting water flow. Slip occurs sharply at the sliding interface, giving rise to high friction that is independent of both sliding velocity and humidity. After $RH_g$, the polymer is in a rubbery state, and a sudden shift in chain relaxation dynamics enables water to flow within the brush layer, creating a fluid contact with a hydrodynamic/fluid-like sliding mechanism to drastically reduce friction and create velocity dependence on the observed shear forces. Since the glass transition occurs across a relatively narrow

range of humidity, the friction behavior switches suddenly. This proposed model holds true both for a grafted brush as well as an adsorbed polymer layer for a given polymer–solvent vapor system.

It must be pointed out that the water in these highly hygroscopic polymers plays a more complex role—one that is not just a property of a polymer undergoing a glass transition. For example, it is neither the case that any polymer film should show sharply switching friction in response to being heated past its $T_g$ nor that any polymer existing in a rubbery state should provide ultralow friction [70]. The large amount of water that a polyelectrolyte absorbs in humid vapors is clearly connected with its ability to display ultralow friction. However, this ultralow friction cannot be realized when the polymer is in a glassy state, regardless of how much water is absorbed. Future experiments should explore if similar properties are observed for non-charged polymers in the presence of organic vapors.

With this understanding, new applications can now be considered that take advantage of surfaces with switchable friction in air environments. Surfaces with such a sharp on/off mechanism find a wide range of industrial and biomedical applications as mechanical actuators for processes such as reversible sticking [22], actuated grabbing and releasing [71], stimuli-induced locomotion [72] or for use as coatings for catheters [21]. Moreover, such devices could be fabricated for both nano- and macro-scale systems and would not require the presence of any liquid solvent or mechanical stimuli. Tunability of ~15–20% RH for the transition was achieved in this study, but we believe a larger range may be achievable to tune the transition for the desired application. The primary phenomenon demonstrated in this work could also be extended to areas outside of switchable friction if future studies could show similarly switching properties above the glass transition such as adhesion, antifouling, or ion diffusion.

## ACKNOWLEDGEMENTS

We acknowledge the National Science Foundation (DMR-2208464) for financial support. AD also acknowledges the financial support from the Knight Foundation (W. Gerald Austen Endowed Chair). SM acknowledges the Goodyear Tire & Rubber Company for financial support under a graduate research intern program (GRIP). We further thank E. Laughlin for the fabrication of custom instruments and other experimental devices, U. Patil, B. Gaire, and N. Orndorf for numerous helpful comments and suggestions, N. Orndorf for assistance with SFG experiments preliminary to those reported here, and A. Patil for assistance with ellipsometry measurements.

## APPENDIX: MATERIALS AND METHODS

### 1. Materials

[2-(methacryloyloxy)ethyl]trimethylammonium chloride (METAC; 75 wt% solution in $H_2O$), 3-sulfopropyl methacrylate potassium salt (SPMA; 98%), ethyl α-bromoisobutyrate (EBiB; 98%), and 2,2'-bipyridyl (bipy; >99%) were purchased from Sigma-Aldrich (St. Louis, USA). L-ascorbic acid sodium salt (99%) was purchased from Alfa Aesar (Haverhill, USA), 3-(triethoxysilyl)propyl 2-bromo-2-methylpropanoate (>93%) was purchased from TCI (Tokyo, Japan), and copper(II) bromide ($CuBr_2$; 99.8%) was purchased from Chem-Impex International (Wood Dale, USA). Ultrapure water (18.2 MΩ.cm, pH 6–7) was obtained from a Millipore filtration system. Additional chemicals of high purity were obtained from common commercial suppliers. All chemicals were used as received without additional purification.

## 2. Synthesis of Brush Surfaces

The substrates used included single-sided polished silicon (1 0 0 orientation) wafers (Silicon Inc., Boise, USA) cut to an approximate size of 2 cm × 2 cm, silicon multi-bounce ATR crystals (Pike Technologies, Madison, USA), and equilateral sapphire prisms (15 × 15 × 15 × 10 mm, Meller Optics, Providence, USA). A schematic of the synthesis procedure is provided in Fig. S13 in the supplementary material. All substrates were cleaned using piranha solution (three parts of a 30% hydrogen peroxide solution mixed with seven parts of sulfuric acid) followed by rinsing and sonication in several changes of ultrapure water. Piranha solution is highly corrosive, and great care must be taken when using it. The substrates were dried using $N_2$ flow prior to plasma treatment for 5 min. (Harrick Plasma, PDC-001-HP) immediately before use. The clean substrates were then placed inside a clean glass 1-L vacuum desiccator (free of any grease) alongside a small glass beaker containing 5 µL of the initiator molecule 3-(triethoxysilyl)propyl 2-bromo-2-methylpropanoate. The desiccator was pumped to a vacuum pressure of < 140 mTorr before being isolated from the vacuum, sealed, and placed in an oven at 80 °C for 4 hr to allow the initiator to covalently deposit onto the substrates via a vapor phase process. The substrates were then removed and sonicated in toluene and ethanol for 10 min. each before drying under $N_2$ flow. Initiator-modified substrates exhibited a static water contact angle of 71 ± 4° with a 1.2-nm initiator layer thickness, as measured by spectroscopic ellipsometry.

After modification of the substrates with an initiator, the surface-initiated activators regenerated by electron transfer atom-transfer radical polymerization (SI-ARGET ATRP) procedure adapted from Zhu and Edmondson [73] was used to grow PMETA and PSPMA brushes. In a typical reaction to prepare PMETA brushes, stock 75 wt% METAC solution (11.04 g total, 39.8 mmol METAC), $CuBr_2$ (1.86 mg, 8.32 µmol), bipy (13.26 mg, 85 µmol), and ascorbic acid (14.9 mg, 85 µmol) were dissolved in 10 mL of a methanol/water mixture [1:1 (v/v)] under constant but gentle $N_2$ flow and allowed to purge for 5 min. In separate sealed vessels, initiator-modified substrates were placed and purged with $N_2$ flow for 10 min. The reaction solution was then transferred via glass syringe to the vessels containing the substrates before being tightly sealed and allowed to react at room temperature for 1 hr., or 12 hours for thicker brushes. When preparing bulk PMETA, the free initiator EBiB (13.5 mg, 70 µmol) was also injected at the start of the reaction under constant stirring. The substrates were then thoroughly rinsed with several changes of methanol and ultrapure water, followed by brief sonication in methanol and ultrapure water for 2 min. each. Bulk PMETA was isolated by shock precipitation into cold dimethylformamide. To prepare the PSPMA brushes, SPMA (8 g, 32.4 mmol), $CuBr_2$ (1.44 mg, 6.44 µmol), bipy (10.16 mg, 65 µmol), and ascorbic acid (11.44 mg, 65 µmol) were dissolved in 20 mL of methanol/water [1:1 (v/v)] to prepare the reaction mixture, and polymerization and cleaning was conducted as described for PMETA but with a 30-min. reaction time. The PMETA-/PSPMA-grafted substrates were finally submerged in a 0.1-M aqueous salt solution of the target counterion (NaCl, NaI, KCl, or LiCl) for 8 min. before being thoroughly rinsed with copious amounts of ultrapure water, blow-dried with $N_2$ flow, and vacuum-dried for 15 min. PMETA-Cl and PSPMA-K brush surfaces showed a thickness of ~110 nm for shorter reaction times and ~350 nm for longer reaction times, as measured by ellipsometry under ~0% controlled relative humidity. The thicker brushes were used for nanoindentation, while the thinner brushes were used for all other experiments. The grafting density was estimated to be around 0.20–0.25 chains/nm$^2$ (supplementary text).

## 3. Generation of Controlled Humidity Vapors

In each of the experiments that follow, vapors of controlled relative humidity (RH) were generated using a custom-built apparatus described in a previous study [74]. A stream of dry $N_2$ from a compressed cylinder (< 0.1% RH) was split into two paths. The stream in one was bubbled through a column of water to produce saturated humid vapors and was allowed to mix with the second path, which served as a stream of dry $N_2$. The relative flow rates of these two streams were adjusted to vary the resulting RH, which was collected and measured downstream in a small vessel fitted with a hygrometer. Using this method, the humidity could generally be controlled within the range <0.1% to ~85–95% RH depending on the experiment. The outlet flow from this vessel was then used for each subsequent experiment.

## 4. Spectroscopic Ellipsometry

Measurements of film refractive index and thickness as a function of RH were performed on brush-grafted silicon wafer substrates using an Accurion EP4 spectroscopic imaging ellipsometer with a liquid cell fit with windows placed for an incident angle of 60°. The cell included inlet and outlet flow ports to allow for continuous flow of humidity-controlled vapors, which could generally be controlled between <0.1% and ~85% RH. For each measurement, values of $\Psi$ and $\Delta$ were recorded between 400 and 1100 nm. The collected data were fit using Accurion EP4Model software with a multi-layer model consisting of a silicon substrate (with optical properties specified by the software), an $SiO_2$ layer (with a refractive index of 1.46 and a thickness of 2.1 nm), an initiator layer (with a refractive index of 1.46 and a thickness of 1.2 nm), a brush layer of unknown refractive index and thickness, and ambient air (with a refractive index of 1.00). The brush was assumed to have a refractive index that does not vary based on wavelength. From this fit, the values of brush refractive index and thickness were extracted. Repeated measurements across different samples were collected, and the resulting average values are reported.

## 5. Friction Measurement

Frictional shear forces were measured using a custom-built setup described in a previous study [66]. Briefly, two pairs of double cantilever springs with attached capacitor plates opposite fixed capacitor plates were used as sensors for normal and shear forces. As the springs deflect due to an external force, a change in capacitance was measured. A calibration using standard weights was performed for each set of springs to convert values of capacitance change to force in the approximate range of 0–100 mN. Two separate motors were used to produce normal loading and lateral shearing motions. The entire friction setup was placed inside a container and was connected to the outlet from the humidity apparatus to produce a controlled humidity environment. Since the container was relatively large, the humidity was allowed to equilibrate for at least 10 min. after reaching a stable humidity, as measured by a hygrometer. Humidity could generally be controlled within the range <0.1% to ~85% RH, and the insertion of a water reservoir allowed for higher humidity ~95–99% RH. While humidity slightly affects the capacitance for the force sensors, the slope between capacitance and spring deflection remains unchanged and, thus, the same calibration can be used (Fig. S14 of the supplementary material).

In each experiment, a hemispherical PDMS lens (prepared according to previously reported procedures [75]) of modulus 0.7 MPa with radius of curvature 2.4 mm was brought into contact with a brush-grafted silicon wafer substrate to a normal load of 30 mN (~60 kPa contact pressure based on observed contact area). After dwelling for 5 min., the lens was slid across the brush substrate at a chosen sliding velocity in one direction while the normal and shear forces were recorded. After the shear force reached a stable value, the average shear force over a period greater than 30 seconds was recorded, which corresponds to the reported values. Repeated measurements across samples were collected, and the resulting average values are reported.

## 6. Attenuated Total Reflectance Spectroscopy

Attenuated total reflectance infrared (ATR-IR) absorbance spectra were collected using a Thermo Fisher Nicolet iS50 Fourier-transform infrared spectrometer with a mercury-cadmium-tellurium detector. An HATR accessory from Pike Technologies was used to collect multibounce ATR spectra of 10 bounces at a 45° angle of incidence in the spectral range of 4000–1500 $cm^{-1}$ with a resolution of 4 $cm^{-1}$, averaged over 32 scans. A silicon crystal with a surface-grafted brush was used as the ATR crystal. The crystal surface was sealed inside a fluid cell fitted with inlet and outlet flow ports, which were used to regulate the flow of humidity-controlled vapors. Spectra were recorded varying the humidity between <0.1% and ~85% RH, after allowing time for the system to reach an equilibrium state in between the changes in humidity. The spectrum of a bare crystal prior to surface treatment was taken as the background spectrum. Automatic baseline subtraction was performed for each collected spectrum using OMNIC data collection and processing software from Thermo Fisher. The reported spectra are representative of the trends observed through repeated measurements across all samples.

## 7. Nanoindentation

Nanoindentation experiments were performed using a Bruker Hysitron TI Premier with a Berkovich 100 nm diamond tip. Instrument compliance and tip area function were calibrated on fused quartz. Displacement-controlled experiments were performed using a peak displacement of 50 nm at 10 nm/s and no holding time to minimize drift. Drift correction was performed before each measurement. The instrument was contained within an environmental chamber which allowed for locally controlled relative humidity. The transducer was re-calibrated after each change of humidity. The values of reduced modulus were extracted from fitting the unloading curves using Hysitron TriboScan software. For these measurements, thicker brushes (~350 nm) were used to reduce the influence of the substrate modulus.

## 8. Sum Frequency Generation Spectroscopy

Sum frequency generation (SFG) spectroscopy utilizes a nonlinear optical process to acquire interface-specific spectral information. An SFG signal may be generated when two beams (one visible and one infrared) overlap at an interface both spatially and temporally. This process is resonantly enhanced when the wavelength of the IR beam corresponds to a vibrational frequency for a chemical group present at the interface. By tuning the wavelength of the IR beam, a spectrum is generated that contains information about the chemical makeup of the interface. In SFG spectroscopy, no signal can generally be produced in a centrosymmetric bulk medium. Details of the SFG system used in this work have been reported previously [76]. In brief, an 800-nm visible and infrared beam tunable 2000–3850 cm$^{-1}$ were produced from a picosecond Spectra Physics laser system with a 1-kHz repetition rate. A sapphire prism with a surface-grafted PMETA-Cl brush was secured into a clean stainless-steel cell and was pressed firmly against a Teflon spacer to create a seal. The cell was constructed with an inlet and outlet port, which were used to maintain a continuous flow of humidity-controlled vapor between <0.1% and ~95% RH and a translating arm with a fixed PDMS lens on one end. The visible and IR beams overlapped at the PMETA-Cl brush–air interface or brush–PDMS interface after the lens is brought into contact through the sapphire prism to produce the SFG beam in a total internal reflection geometry whereby the incident angles were chosen as the critical angle for the given interface. The incident angles for the IR and visible beams—which were 42° and 40.5°, respectively, for the brush–air interface and 8° and 6.5°, respectively, for the brush–PDMS interface—were measured with reference to the normal of the sapphire prism. PDMS contact was achieved with a load of ~800 kPa (as estimated from the contact area) to ensure the laser beam spot would fall within a sufficiently large contact zone. SFG spectra were collected in both PPP and SSP polarization combinations, where the letters (in sequential order) correspond to the polarization states of the SFG, visible, and IR beams; P-polarized light corresponds to an electric field perpendicular to the surface plane, while S-polarized light corresponds to an electric field parallel to the surface plane. The reported spectra are representative of the trends observed through repeated measurement across samples.